\documentclass[aps,prd,preprint,superscriptaddress,tightenlines,nofootinbib,showpacs]{revtex4}

\usepackage{graphicx}
\usepackage{dcolumn}
\usepackage{bm}

\newcommand{\gamgam}{\gamma\gamma}
\newcommand{\jpipi}{\pi^{+}\pi^{-}J/\psi}

\newcommand{\deltm}{\mathrm{M}(\pi^{+}\pi^{-}l^{+}l^{-}) - 
		    \mathrm{M}(l^{+}l^{-})}

\newcommand{\jll}{J/\psi \rightarrow l^{+}l^{-}}
\newcommand{\jee}{J/\psi \rightarrow e^{+}e^{-}}
\newcommand{\jmm}{J/\psi \rightarrow \mu^{+}\mu^{-}}

\newcommand{\Gammagg}{\Gamma_{\gamgam}(X(3872))}
\newcommand{\Gammaee}{\Gamma_{ee}(X(3872))}

\newcommand{\xBR}{{\cal{B}}(X\rightarrow\jpipi)}

\newcommand{\xGGUL}{(2J+1)\Gammagg\xBR}
\newcommand{\xISRUL}{\Gammaee\xBR}

\begin{document}

\preprint{CLNS 04-1892}       
\preprint{CLEO 04-12}         

\title{Search for $X$(3872) in $\gamgam$ Fusion and Radiative Production at CLEO}


\author{S.~Dobbs}
\author{Z.~Metreveli}
\author{K.~K.~Seth}
\author{A.~Tomaradze}
\author{P.~Zweber}
\affiliation{Northwestern University, Evanston, Illinois 60208}
\author{J.~Ernst}
\author{A.~H.~Mahmood}
\affiliation{State University of New York at Albany, Albany, New York 12222}
\author{H.~Severini}
\affiliation{University of Oklahoma, Norman, Oklahoma 73019}
\author{D.~M.~Asner}
\author{S.~A.~Dytman}
\author{W.~Love}
\author{S.~Mehrabyan}
\author{J.~A.~Mueller}
\author{V.~Savinov}
\affiliation{University of Pittsburgh, Pittsburgh, Pennsylvania 15260}
% ...
\collaboration{CLEO Collaboration} 
\noaffiliation


\date{January 27, 2005}

\begin{abstract} 
We report on a search for the recently reported $X$(3872) state using 
15.1 fb$^{-1}$ of $e^{+}e^{-}$ data taken in the $\sqrt{s}$ = 9.46-11.30 GeV 
region.  Separate searches for the production of the $X$(3872) in 
untagged $\gamgam$ fusion and $e^{+}e^{-}$ annihilation following 
initial state radiation are made by taking advantage of the unique 
angular correlation between the leptons from the decay $\jll$ in $X$(3872) 
decay to $\jpipi$.  No signals are observed in either case, 
and 90$\%$ confidence upper limits are established as 
$\xGGUL$ $<$ 12.9 eV and $\xISRUL$ $<$ 8.3 eV.
\end{abstract}

\pacs{12.39.Mk,13.66.Lm,13.25.Gv,14.40.Gx}
\maketitle

The Belle Collaboration recently reported the observation of a narrow 
state, $X$(3872), in the decay 
{$B^{\pm}$$\rightarrow$$K^{\pm}X$}, $X$ $\rightarrow$ $\jpipi$, $\jll$ 
($l$ = $e,\mu$) \cite{xBELLE}.  
The observation was confirmed by the CDF II \cite{xCDFII}, D{\O} \cite{xD0},
and {\slshape{B{\scriptsize{A}}B{\scriptsize{AR}}}} \cite{xBABAR} 
collaborations, with consistent results, M($X$) = 3872 $\pm$ 1 MeV/c$^{2}$, 
and $\Gamma(X)$ $\le$ 3  MeV/c$^{2}$.

Many different theoretical interpretations of the nature of the $X$(3872) 
state and its possible quantum numbers have been proposed 
\cite{xccbarbound1,xccbarbound2,xmole1,xmole2,xexotic,xthref1,xthref2,xthref3,xthref4,xthref5,xglueball}.  These include that (a) $X$(3872) is a 
charmonium state \cite{xccbarbound1,xccbarbound2}; (b)  $X$(3872) is a 
$D^{0}\bar{D}$$^{*0}$ loosely bound ``molecular'' 
state \cite{xmole1,xmole2} since its mass 
is close to (M$_{D^{0}}$ + M$_{\bar{D}^{*0}}$) = 3871.3 $\pm$ 1.0 
MeV/c$^{2}$ \cite{2004partlist}; and
(c) $X$(3872) is an exotic state \cite{xexotic}. 

No positive signals for $X$(3872) have been observed in searches for the decay 
channels $X$(3872) $\rightarrow$ $\gamma\chi_{c1}$ \cite{xBELLE}, 
$\gamma\chi_{c2}$, $\gamma J/\psi$, 
$\pi^{0} \pi^{0} J/\psi$ \cite{xOmegaJpsiBELLE}, 
$\eta J/\psi$ \cite{xJpsiEtaBABAR}, $D^{+}D^{-}$, $D^{0}\bar{D}$$^{0}$, 
and $D^{0}\bar{D}$$^{0}\pi^{0}$ \cite{xDDBELLE}, or for possible charged 
partners of $X$(3872) \cite{xChargedBABAR}.  Yuan, Mo, and 
Wang \cite{xISRBES} have used 
22.3 pb$^{-1}$ of BES data at $\sqrt{s}$ = 4.03 GeV to determine the 
upper limit of $\xISRUL$ $<$ 10 eV (90$\%$ C.L.) for $X$(3872) production 
via initial state radiation (ISR).  Belle \cite{xOmegaJpsiBELLE} has 
recently reported a small enhancement in the $\pi^{+}\pi^{-}\pi^{0}J/\psi$ 
effective mass near the $X$(3872) mass.  

The variety of possibilities for the structure of $X$(3872) suggests that,
irrespective of the models, it is useful to limit the $J^{PC}$ of
$X$(3872) as much as possible.  The present investigation is designated to
provide experimental constraints for the $J^{PC}$ of  $X$(3872) by studying
its production in $\gamma\gamma$ fusion and ISR, and its decay into
$\pi^{+}\pi^{-}J/\psi$. Production of $X$(3872) in $\gamma\gamma$ fusion
can shed light on the positive charge parity candidate states,
charmonium states 2$^{3}$P$_{0}$, 2$^{3}$P$_{2}$ and
1$^{1}$D$_{2}$ \cite{xccbarbound1,xccbarbound2}, and the $0^{-+}$ molecular
state \cite{xmole1,xmole2}. ISR   production can address the $1^{--}$ vector 
state.
 
The data used for this $X$(3872) search were collected at the Cornell Electron 
Storage Ring (CESR) with the detector in the CLEO III 
configuration \cite{CLEOIIIDetector}.  
The detector is cylindrically symmetric and 
provides 93$\%$ coverage of solid 
angle for charged and neutral particle 
identification.  The detector components important for this analysis are 
the drift chamber (DR), CsI crystal calorimeter (CC), and muon identification 
system (MIS). The DR and CC are operated within a 1.5 T magnetic field 
produced by a superconducting solenoid located directly outside of the CC.   
The DR detects charged particles and measures their momenta and ionization 
energy loss (dE/dx).  The CC allows precision measurements of 
electromagnetic shower energy and position.  The MIS consists of proportional 
chambers placed between layers of the magnetic field return iron to detect 
charged particles which penetrate a minimum of three nuclear 
interaction lengths.

The data consist of a 15.1 fb$^{-1}$ sample of $e^{+}e^{-}$ collisions at or 
near the energies of the $\Upsilon(nS)$ resonances ($n=1$--$5$), and in the 
vicinity of the $\Lambda_{b}\bar{\Lambda}_{b}$ threshold.  Table 
\ref{tab:infotable} lists the six different initial center-of-mass 
energies and integrated luminosities of the data samples.  

\begin{table*}[ht]
\caption{Data samples and MC determined detection efficiencies used for the 
present $X$(3872) search.  $\langle\sqrt{s_{i}}$ $\rangle$ are the average 
center-of-mass energies of $\Upsilon$($1S-5S$) and 
$\Lambda_{b}\bar{\Lambda}_{b}$ threshold measurement and 
${\cal{L}}_{i}$($e^{+}e^{-}$) is the $e^{+}e^{-}$ integrated luminosity at 
$\sqrt{s_{i}}$. 
The efficiencies $\epsilon_{\gamgam,i}$ and $\epsilon_{ISR,i}$ are the sums 
of the efficiencies $\epsilon_{ee,i}$ and $\epsilon_{\mu\mu,i}$ for 
electron and muon detection, respectively.  The $\gamgam$ fusion/ISR 
separation, as described in the text, is applied to the respective MC samples.}
\begin{center}
\begin{tabular}{|c|c|c||c|c|c||c|c|c|} 
\hline
& $\langle\sqrt{s_{i}}$ $\rangle$ & ${\cal{L}}_{i}$($e^{+}e^{-}$) 
& \multicolumn{3}{|c||}{$\gamgam$ Fusion} 
& \multicolumn{3}{|c|}{ISR}\\
\hline
&(GeV) & (fb$^{-1}$) &
$\epsilon_{ee,i}$ & $\epsilon_{\mu\mu,i}$ & $\epsilon_{\gamgam,i}$ &
$\epsilon_{ee,i}$ & $\epsilon_{\mu\mu,i}$ & $\epsilon_{ISR,i}$ \\
\hline
$\Upsilon(1S)$ & 9.458  & 1.47 & 0.128(4) & 0.160(4) & 0.288(6) &
0.065(3) & 0.083(3) & 0.148(4) \\
$\Upsilon(2S)$ & 10.018 & 1.84 & 0.121(3) & 0.151(4) & 0.272(5) & 
0.054(2) & 0.062(3) & 0.116(4) \\
$\Upsilon(3S)$ & 10.356 & 1.67 & 0.115(3) & 0.137(4) & 0.252(5) & 
0.042(2) & 0.043(2) & 0.085(4) \\
$\Upsilon(4S)$ & 10.566 & 8.97 & 0.123(4) & 0.145(4) & 0.268(6) & 
0.0186(14) & 0.0165(13) & 0.0351(19) \\
$\Upsilon(5S)$ & 10.868 & 0.43 & 0.113(3) & 0.139(4) & 0.252(5) &
0.0025(5) & 0 & 0.0025(5) \\
$\Lambda_{b}\bar{\Lambda}_{b}$ threshold & 11.296 & 0.72 & 
0.104(3) & 0.126(4) & 0.230(5) &
0.0001(1) & 0 & 0.0001(1) \\
\hline
\end{tabular}
\end{center}
\label{tab:infotable}
\end{table*}

Resonance production by untagged $\gamgam$ fusion and by ISR
have similar characteristics. The undetected electrons in untagged
 $\gamgam$ fusion and the undetected radiated photons in ISR
 have angular distributions sharply peaked along the beam axis. Both 
processes have total observed energy (E$_{tot}$) much smaller than
the center-of-mass energy, $\sqrt{s}$, of the original $e^{+}e^{-}$ 
system, and have small observed transverse momentum. 
The detailed characteristics for $\gamgam$ fusion and ISR resonance 
production are studied by generating signal Monte Carlo (MC) samples using 
GEANT 3.21/11 \cite{GEANTMC} to simulate the CLEO III detector. For $X$(3872) 
production by $\gamgam$ fusion the formalism of Budnev {\itshape{et al.}} 
\cite{ggcs} is used. For ISR resonance production the formalism 
of M. Benayoun {\itshape{et al.}} \cite{isrprod} is used.

A fully reconstructed event has four charged particles and zero net 
charge.  All charged particles must lie within the drift chamber volume and 
satisfy standard requirements for track quality and distance of closest 
approach to the interaction point.  Events must also have detected 
E$_{tot}$ $<$ 6 GeV.  The $X$(3872) resonance corresponds to $\Delta$M 
$\equiv$ $\deltm$ = 0.775 GeV/c$^{2}$, and we designate $\Delta$M = 
0.63--0.7 and 0.85--0.92 GeV/c$^{2}$ as background regions.  
Signal-to-background studies are performed to optimize signal efficiency 
and background suppression.  Selection criteria optimized the efficiency 
for reconstructing $\gamgam$ fusion MC events.  The selection variables 
optimized are the total neutral energy (E$_{neu}$) of the event, 
total transverse momentum of the four charged tracks (p$_{tr}$), 
lepton pair invariant mass (M($l^{+}l^{-}$)) of the $\jll$ decay, 
and particle identification (PID) of the charged tracks.  
Based on the optimization studies, events are selected 
with E$_{neu}$ $<$ 0.4 GeV and p$_{tr}$ $<$ 0.3 GeV/c.  Events with a 
$\jee$ decay require both electron candidates to satisfy dE/dx and shower 
energy criteria consistent with the electron hypothesis, and to have 
invariant mass in the range M($e^{+}e^{-}$) = 2.96-3.125 GeV/c$^{2}$.  
Events with a $\jmm$ decay require both muon candidates to appear as minimum 
ionizing particles in the CC, with at least one muon penetrating the number of 
interaction lengths in the MIS consistent with its momentum, and to have 
invariant mass in the range M($\mu^{+}\mu^{-}$) = 3.05-3.125 GeV/c$^{2}$.  
Each of the two pions recoiling against the $J/\psi$ is required to satisfy 
the dE/dx pion hypothesis.

\begin{figure}
\includegraphics*[width=3.2in]{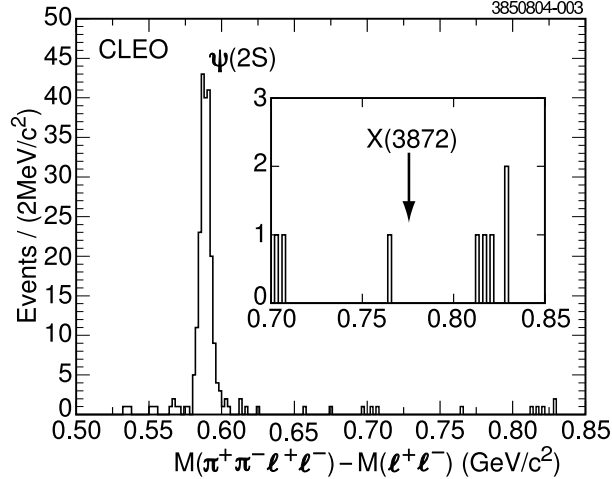}
\caption{Data events as function of $\Delta$M $\equiv$ $\deltm$.  The 
$\psi$($2S$) is clearly visible and no apparent enhancement 
is seen in the $X$(3872) region.}
\label{fig:alldm}
\end{figure}

Figure \ref{fig:alldm} shows the $\Delta$M distribution for data events 
which pass the selection criteria and have $\Delta$M = 
0.514--0.850 GeV/c$^{2}$.  A $\psi$($2S$) signal 
is clearly visible while no enhancement is apparent in the $X$(3872) region.  
The observed number of $\psi$($2S$) events is determined by fitting 
the $\psi$($2S$) region with a mass-independent background and 
a resonance whose shape is determined by fitting the $\psi$($2S$) peak in 
the ISR MC simulation.  
The observed number of $\psi$($2S$) is $N_{ISR}(\psi(2S))$ = 206 $\pm$ 15 
events.  A MC simulation predicts $N_{ISR}(\psi(2S))$ = 226 $\pm$ 11 events.

\begin{figure}
\includegraphics*[width=3.4in]{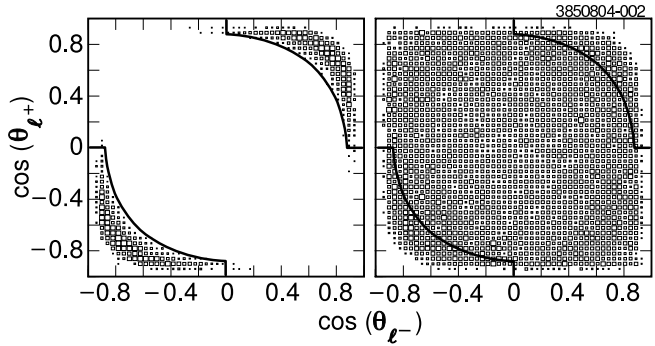}
\caption{MC predictions for the two-dimensional lepton pair cos($\theta$) 
distributions for the $X$(3872): ISR (left) and $\gamgam$ fusion (right).  The 
lines indicate how the ISR resonance and $\gamgam$ fusion samples 
are separated.}
\label{fig:2dcsth}
\end{figure}

At $\sqrt{s}$ $\sim$ 10 GeV, a feature unique to the ISR mediated 
production of a vector resonance which decays via $\jpipi$, $\jll$ is the 
correlation between the angles $\theta_{l+}$ and $\theta_{l-}$ in the 
laboratory system.
Figure \ref{fig:2dcsth} shows the MC prediction for the two-dimensional 
cos($\theta$) distributions for leptons from $X$(3872) decay for the 
ISR mediated and $\gamgam$ fusion productions.  As shown in Figure 
\ref{fig:2dcsth}, a parabolic cut applied to the two-dimensional 
cos($\theta$) distribution efficiently separates the events from the 
two production processes.  With this cut, the $\gamgam$ region contains a 
0.6$\%$ contamination from ISR production, and the ISR sample contains a 
14$\%$ contamination from $\gamgam$ fusion production if we assume for 
an illustrative purpose that $(2J+1)\Gamma_{\gamgam}(X)$ = 
$\Gamma_{ee}(X)$.  Here $J$ is the total spin and $\Gamma_{\gamgam}$ 
($\Gamma_{ee}$) is the two-photon ($e^{+}e^{-}$) partial width of $X$(3872). 

The efficiencies as determined by MC simulations of $X$(3872) production and 
decay following $\gamgam$ fusion and ISR are listed in 
Table \ref{tab:infotable}.  The $X$(3872) and $J/\psi$ are decayed 
according to phase space in the MC simulations.  The same selection 
criteria are applied to both MC 
samples except for the lepton pair cos($\theta$) cut described above.

\begin{figure}
\includegraphics*[width=3.0in]{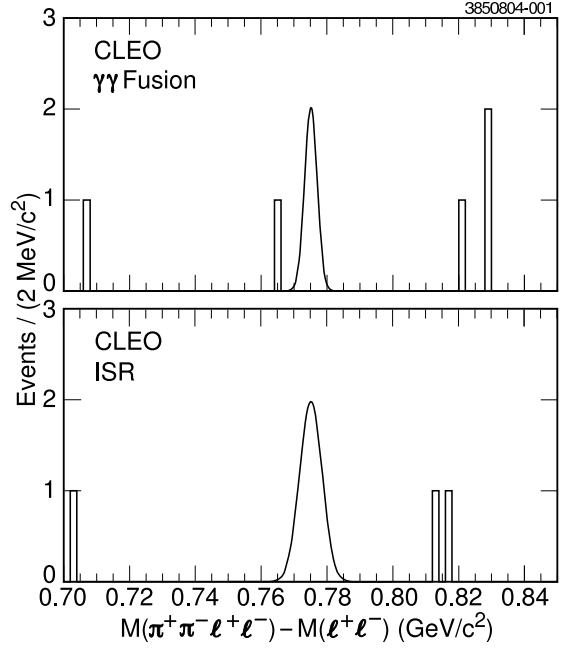}
\caption{Distributions of data events as function of $\Delta$M $\equiv$ 
$\deltm$ for $\gamgam$ fusion (top) and ISR (bottom) events in the 
region $\Delta$M = 0.7--0.85 GeV/c$^{2}$.  The mass resolution functions 
determined from MC simulations are shown on an arbitrary scale 
at $\Delta$M = 0.775 GeV/c$^{2}$.}
\label{fig:inddm}
\end{figure}

The separate $\Delta$M distributions for the data in the $X$(3872) search 
region for $\gamgam$ fusion and ISR mediated resonance production are shown in 
Figure \ref{fig:inddm}.  The number of observed $X$(3872) events 
($N_{\gamgam,ISR}(X(3872))$) is determined by maximum likelihood fits of 
the $\Delta$M data using mass-independent backgrounds and the appropriate 
detector resolution functions for the two production processes.  The 
detector resolution functions are determined by the MC simulations fitted 
with double Gaussians which are illustrated in 
Figure \ref{fig:inddm}.  The 90$\%$ confidence upper limits on the 
observed number of $X$(3872) events in untagged $\gamgam$ fusion and ISR 
mediated resonance production are determined to be $N_{\gamgam,ISR}(X(3872))$ 
$<$ 2.36 for both processes.

\begin{figure}
\includegraphics*[width=3.0in]{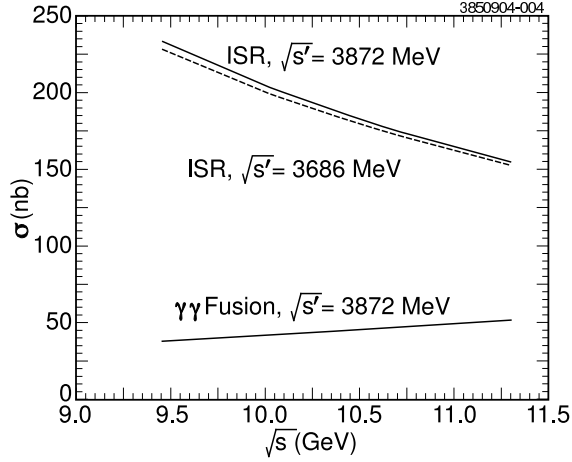}
\caption{Cross sections for $e^{+}e^{-}$ collisions at $\sqrt{s}$ to 
produce reduced CM energy, $\sqrt{s^{\prime}}$, for $\gamgam$ 
fusion \cite{ggcs} with $\sqrt{s^{\prime}}$ = 3872 MeV 
and ISR \cite{isrprod} with $\sqrt{s^{\prime}}$ = 3872 MeV and 
$\sqrt{s^{\prime}}$ = 3686 MeV.}
\label{fig:crbehave}
\end{figure}

The cross section for $\gamgam$ fusion or ISR mediated production
of the $X$(3872) resonance with total angular momentum $J$, and decay
through $\jpipi$, $\jll$, is
\begin{displaymath}
(\frac{\mathrm{C}_{\gamgam,ISR}}
{\mathrm{M}})(2J+1)\Gamma_{\gamgam,ee}(X)\xBR = 
\end{displaymath}
\begin{equation}
\frac{N_{\gamgam,ISR}(X(3872))}{{\cal{B}}(\jll)\sum_{i} 
\mathcal{L}_{i}(e^{+}e^{-}) \epsilon_{\gamgam,ISR,i} 
\sigma(\sqrt{s_{i}})_{\gamgam,ISR}} 
\end{equation}
where C$_{\gamgam,ISR}$ are constants, M = 3872 MeV/c$^{2}$, $\sqrt{s_{i}}$, 
$\mathcal{L}_{i}$($e^{+}e^{-}$), and $\epsilon_{\gamgam,ISR,i}$ are as 
listed in Table \ref{tab:infotable}, and 
$\sigma$($\sqrt{s_{i}}$)$_{\gamgam,ISR}$ are as shown in 
Figure \ref{fig:crbehave}.  The branching fraction ${\cal{B}}(\jll)$ 
= (5.91$\pm$0.07)$\%$ is the average PDG branching fraction of $\jee$ and 
$\jmm$ \cite{2004partlist}. 
This leads to the 90$\%$ confidence upper limits 
\begin{displaymath}
\xGGUL < 10.9~\mathrm{eV}
\end{displaymath}
for $X$(3872) production in $\gamgam$ fusion, and
\begin{displaymath}
\xISRUL < 7.3~\mathrm{eV}
\end{displaymath}
for $X$(3872) production via ISR.

Systematic uncertainty in the above limits arises from 
possible biases in the detection efficiency and estimated background level.  
These are studied by varying the track quality, $\gamgam$ 
fusion/ISR separation, and selection criterion optimized in the 
signal-to-background studies.  Other systematic uncertainties are 
from the $e^{+}e^{-}$ luminosity measurement and $\jll$ branching 
fractions.  Adding these in quadrature, the
total systematic uncertainties
in  $\gamgam$ fusion and ISR are 18.5$\%$ and 13.2$\%$, respectively.
A conservative way to incorporate these systematic uncertainties 
is to increase the measured upper limits by these amounts.  This leads to 
the 90$\%$ confidence upper limits
\begin{displaymath}
\xGGUL < 12.9~\mathrm{eV}
\end{displaymath}
for $X$(3872) which has positive C parity, and
\begin{displaymath}
\xISRUL < 8.3~\mathrm{eV}
\end{displaymath} 
for $X$(3872) being a vector meson with $J^{PC}$ = 1$^{--}$.

If $\cal{B}$($B^{\pm}$$\rightarrow$$K^{\pm}X(3872)$) $\approx$ 
$\cal{B}$($B^{\pm}$$\rightarrow$$K^{\pm}\psi(2S))$ = 
(6.8$\pm$0.4)$\times$10$^{-4}$ \cite{2004partlist} is assumed, 
we obtain $\xBR$ $\approx$ 0.02 from both the Belle \cite{xBELLE} and 
{\slshape{B{\scriptsize{A}}B{\scriptsize{AR}}}} \cite{xBABAR} results.  
This leads to 90$\%$ confidence upper limits
\begin{displaymath}
(2J+1)\Gamma_{\gamgam}(X(3872)) < 0.65~\mathrm{keV},
\end{displaymath} 
and 
\begin{displaymath}
\Gamma_{ee}(X(3872)) < 0.42~\mathrm{keV}.
\end{displaymath} 
The (2$J$+1)$\Gamma_{\gamgam}(X(3872))$ upper limit is almost 1/4 
the corresponding values for $\chi_{c0}$ and $\chi_{c2}$, 
but it is nearly 6 times larger than the prediction for 
the 1$^{1}$D$_{2}$ state of charmonium \cite{xccbartwophot}.
The upper limit for $\Gamma_{ee}$($X$(3872)) is comparable to the measured 
electron width of $\psi(3770)$ and is about 1/2 that of $\psi(4040)$.  We 
also note that the ratio 
$N_{ISR}(X(3872))/N_{ISR}(\psi(2S))$ $<$ 0.01 (90$\%$ C.L.). 

We gratefully acknowledge the effort of the CESR staff 
in providing us with excellent luminosity and running conditions.
This work was supported by the National Science Foundation
and the U.S. Department of Energy.

\end{document}